\documentclass[sn-mathphys]{sn-jnl}% Math and Physical Sciences Reference Style

\jyear{2023}%

\theoremstyle{thmstyleone}%
\theoremstyle{thmstyletwo}%

\theoremstyle{thmstylethree}%
\definecolor{Dred}{rgb}{0.312,0.070,0.070}

\renewcommand{\citet}{\citep}
\newcommand{\Number}[1]{\ifnum#1<10\relax0\number#1\else\number#1\fi}
\newcommand{\isodate}{
\count151=\time
\divide\count151 by 60
\count151=\count151
\multiply\count151 by 60
\count152=\time
\advance\count152 by -\count151
\divide\count151 by 60
\count152=\count151
\multiply\count151 by 60
\count153=\time
\advance\count153 by -\count151
\Number{\year}.\Number{\month}.\Number{\day}--\Number{\count152}:\Number{\count153}
}

\newcounter{note}
\let\oldmarginpar\marginpar
\renewcommand\marginpar[1]{\-\oldmarginpar[\raggedleft\footnotesize #1]%
{\raggedright\footnotesize #1}}

\begin{document}

\title[Precise VLBI/GNSS ties with micro-VLBI]{\par\vspace{-6ex}\par Precise VLBI/GNSS ties with micro-VLBI}

\author*[1]{\fnm{Leonid} \sur{Petrov}}\email{Leonid.Petrov@nasa.gov}

\author[2]{\fnm{Johnathan} \sur{York}}\email{york@arlut.utexas.edu}

\author[2]{\fnm{Joe} \sur{Skeens}}\email{jskeens1@utexas.edu}

\author[2]{\fnm{Richard} \sur{Ji-Cathriner}}\email{rcathriner@arlut.utexas.edu}

\author[2]{\fnm{David} \sur{Munton}}\email{dmunton@arlut.utexas.edu}

\author[2]{\fnm{Kyle} \sur{Herrity}}\email{kherrity@arlut.utexas.edu}

\equalcont{\relax}

\affil*[1]{\orgdiv{}, \orgname{NASA Goddard Space Flight Center}, 
\orgaddress{\street{8800}, \city{Greenbelt}, \postcode{20771}, 
\state{MD}, \country{USA}}}

\affil[2]{\orgdiv{}, \orgname{University of Texas at Austin}, 
\orgaddress{\street{10,000 Burnet Rd}, \city{Austin}, \postcode{78758}, 
\state{TX}, \country{USA}}}

\abstract{
          We present here a concept of measuring local ties between collocated
          GNSS and VLBI stations using the microwave technique that effectively
          transforms a GNSS receiver to an element of a VLBI network.
          This is achieved by modifying the signal chain that allows
          to transfer voltage of the GNSS antenna to a digitizer via
          a coaxial cable. We discuss the application of this technique
          to local tie measurement. We have performed observations with
          a GNSS antenna and FD-VLBA radiotelescope and detected a strong 
          interferometric signal from both radiogalaxies and GNSS satellites.
}


\maketitle

%% \vspace{-106ex} \hfill \framebox{\Rdb{\LARGE\bf Draft of \isodate}} \vspace{99ex}   %%

\section{Introduction}

   Space geodetic observation with Very Long Baseline Interferometry 
(VLBI), Global Navigation Satellite System (GNSS), Satellite Laser Ranging 
(SLR), or Doppler Orbitography and Radiopositioning Integrated by Satellite 
(DORIS) involves a measurement of travel time of electromagnetic radiation between 
an emitter and a receiver and/or rates of its change. The position of a ground 
space geodesy instrument is referred to its own unique reference point. 
In a similar way, the position of a space-borne emitter or receiver is referred 
to its own reference point. 

  Space geodesy techniques have their strengths and weaknesses. VLBI provides 
a reference to inertial space, SLR provides a reference to the Earth's center 
of mass, DORIS provides wide spatial coverage, and GNSS is able to 
sense site deformations with a fine time resolution. It was recognized over 
20~years ago that a combination of all space geodesy techniques has 
the potential to provide the most accurate results by mitigating the weaknesses
of each individual technique \citep[see for example,][]{r:alt02}. 
Combination implies that observations necessarily must have something common 
that ties them together. Ties can be direct in the form of a position vector 
between either a space-borne or a ground-based reference point that is 
precisely known, or indirect, for instance in the form of Earth rotation 
parameters that affects all ground stations.

   A number of sites have instruments of more than one technique at 
distances of 30--500 meters. Direct measurement of their positions with 
respect to each other can establish direct ties. Survey 
techniques measure angles and distances {\it between markers}. 
These measurements can reach accuracy of 1--3~mm. However, the ties should
provide the positions of {\it technique reference points}. Reference points 
of microwave techniques, such as VLBI and GNSS, cannot be directly pin-pointed 
by markers. An offset of a reference point with respect to a marker is 
inferred. In the case of GNSS ground stations, an offset between a marker 
on the instrument and its phase center can be calibrated, for instance, 
in an anechoic chamber. In the case of VLBI radiotelescopes, markers are put 
on the antenna, and the position of a geometric reference point on a fixed axis 
that is a projection of a moving axis is derived from processing a cloud 
of points measured with a total station at different antenna azimuths and 
elevations. Then an assertion is made that the geometric reference point 
coincides with the reference point estimated in VLBI data analysis. The 
validity of that assertion cannot be evaluated.

  The positions of microwave reference points provided by data analysis 
of both the GNSS and VLBI techniques may have biases with respect to the geometric
reference points. As long as these biases are permanent and do not depend 
on any other variable parameters, they can remain unnoticed. For a number
of applications, for instance for a study of motions caused by plate 
tectonics or for mean sea level determination, permanent biases are 
irrelevant. \citet{r:sarti11} showed that antenna gravity deformation
caused a 7~mm offset of the microwave reference point of VLBI station 
{\sc medicina}, whose position was determined from analysis of VLBI group 
delay with respect to a geometric refrence point determined from a local survey. 
Mismodeling antenna gravity deformation 
will not affect the least square fit and may not be noticed, but biases 
in local tie measurements comparable or exceeding the internal accuracy 
of GNSS and VLBI techniques make them close to useless.

  The fundamental problem of tie measurements with local surveys is that 
the optical technique used by a survey instrument, such as a total 
station, cannot measure the phase center of a microwave technique. While the 
accuracy of measurements of a vector between markers can be evaluated 
from a scatter of residuals, the accuracy of a vector between a marker 
to a phase center is poorly known. That makes tie vectors determined 
with local surveys unreliable. A typical discrepancy between VLBI 
positions determined from analysis of group delays and GNSS positions 
reduced to the VLBI reference point via tie vectors determined with local 
surveys is 5--20~mm \citep{r:ray05}. Lack of realistic uncertainties of tie 
vectors does not allow us to interpret these discrepancies because we do not 
know whether they are due to systematic errors of space geodetic techniques 
or due to error in tie vectors. That motivated us to seek for alternative 
measurements of tie vectors.

\section{A microwave technique for VLBI/GNSS tie measurements}

  We are leveraging the High Rate Tracking Receiver (HRTR) to serve as both
an advanced software-defined GNSS receiving system and a general purpose 
L-band receiver \citep{r:hrtr12,r:yor14}. It directly digitizes voltage from 
the receiver in a range of 1 to 2~GHz at a rate of 2~gigasamples per second. 
To access a larger extent of the signal in the 1--2~GHz range, we modified 
a commercial GNSS antenna. Specifically, we removed the internal amplifier 
and narrowband bandpass filters that are provided, and we have replaced 
them with an alternate amplification and filtering stage of our own design. 
The aggregate RF system including modified components has a passband of 
approximately [1.10, 1.65]~GHz. The antenna elements are not altered in 
this modification.

  The HRTR performs digital downconversion of the input samples and produces 
up to nine independently chosen frequency bands 40.912~MHz wide. The HRTR 
also allows us to configure the bit depth of the received signal. We utilized 
complex encoding for our work, using one bit for the in-phase 
and one bit for the quadrature voltage. Datastreams with the baseband 
signal from each band are recorded to a general purpose RAID of magnetic 
harddrives. 

  In additional to recording voltage from the receiver, HRTR simultaneously 
computes conventional GNSS observables on civil GNSS signals in real time 
and provides an output convertible to RINEX format. We can recompute 
conventional GNSS observables by processing digital records of the baseband 
signal if needed.

  We noticed that the HRTR has a striking similarity to a radiotelescope
that is an element of a VLBI network. Like a radiotelescope, the HRTR 
digitizes voltage from the antenna and records the data with time stamps 
from a precise clock. HRTR data are processed after the experiment in 
a similar way as VLBI. Extending this analogy further, we came to the idea 
of using a HRTR itself as an element of a VLBI network. A GNSS antenna with
an effective diameter of $\sim\! 0.08$~meter surrounded by 0.38~m wide choke 
ring is roughly four orders of magnitude less sensitive than a 12--30~meter 
radiotelescope, it operates at a lower frequency, and at 
a first glance does not look competitive. However, the use of a GNSS 
antenna as an element of a VLBI network is very promising for local tie 
measurement. Observations at short baselines 30--3000~meters are
almost insensitive to the atmospheric path delay. \citet{r:hase99,r:var21} 
demonstrated that baseline length repeatability at 
a sub-millimeter level has been achieved from processing of phase delays
at short baselines. We should stress that the baseline vector between two 
antennas evaluated from VLBI observations is between the microwave reference 
points of the radiotelescopes. Therefore, by processing GNSS/radiotelescope 
data, we can eliminate the weakest link in the measurement chain of a tie 
vector with the use of conventional local surveys: the offset between a marker 
and a microwave reference point. 

\begin{figure}[h]
  \includegraphics[width=0.99\textwidth]{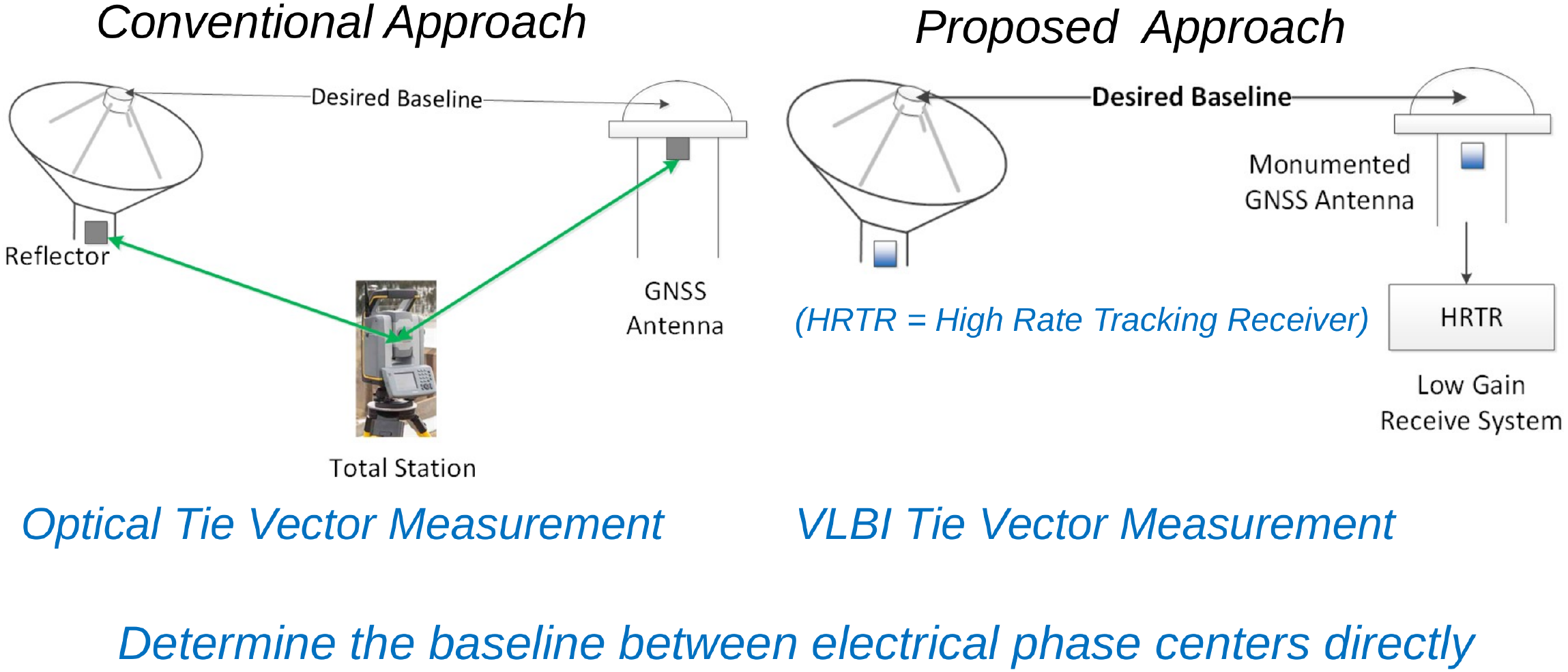}
  \caption{Measurement concept of VLBI/GNSS ties with a microwave technique.}
  \label{f:meas}
%%  \par\vspace{-2ex}\par
\end{figure}

  The measurement concept is presented in Figure~\ref{f:meas}. The voltage of 
the emission received by the GNSS antenna in a range of 1.0 to 2.0~GHz 
is transferred via a coaxial cable to an analog-to-digital converter 
and recorded. The digital records 
are re-sampled and re-coded to VDIF format that is commonly accepted 
in radio astronomy. Emission received by a VLBI antenna is processed 
the same way, but with different hardware. The output of both the GNSS
and VLBI antennas is recorded in the same format. Further
processing is performed exactly the same way as processing of any other 
VLBI data: the data are correlated, the fringe fitting procedure 
finds group delays and phase delay rates that maximize the fringe 
amplitude coherently averaged over time and frequency, and finally group and phase
delays are used for determination of the baseline vector using least 
squares. Therefore, we expect that a position vector of a GNSS 
antenna/radiotelescope baseline will be determined with the same 
sub-millimeter accuracy as position vectors of other short baselines.

\section{Early results}

  Implementation of VLBI with a GNSS antenna requires overcoming a number 
of difficulties (J. Skeens et al. (2023), paper in preparation). It is 
essential that the HRTR does not perform any analog signal transformation. 
It simply digitizes signal as is, performs digital filtering into several bands, 
and writes the digital signal. This early digitization approach shifts the burden 
of signal processing to programming. This facilitates the tuning of the 
processing pipeline since digital records can be re-processed as many times 
as needed.

  We performed three 3~hr long observing sessions in 2022 between 
two transportable HRTRs and a 25~m radiotelescope {\sc fd-vlba}.
That radiotelescope is a part of the Very Long Baseline Array dedicated
for VLBI and has been operating since 1991. It is equipped with an 
H-maser clock. The antenna has a number of very sensitive receivers, including 
the one that operates at L-band. We put the first HRTR within 90~m of 
the {\sc fd-vlba}. That HRTR was stabilized by a Rubidium clock. We put the 
second HRTR within 9000~m of {\sc fd-vlba} near the NASA VLBI station 
{\sc macgo12m}. This HRTR was stabilized by the H maser clock used by {\sc macgo12m}. 
Since {\sc macgo12m} does not have the technical capability to observe below 2~GHz, 
we performed observations at only the two HRTRs and {\sc fd-vlba}.

\begin{figure}[h]
  \includegraphics[width=0.487\textwidth]{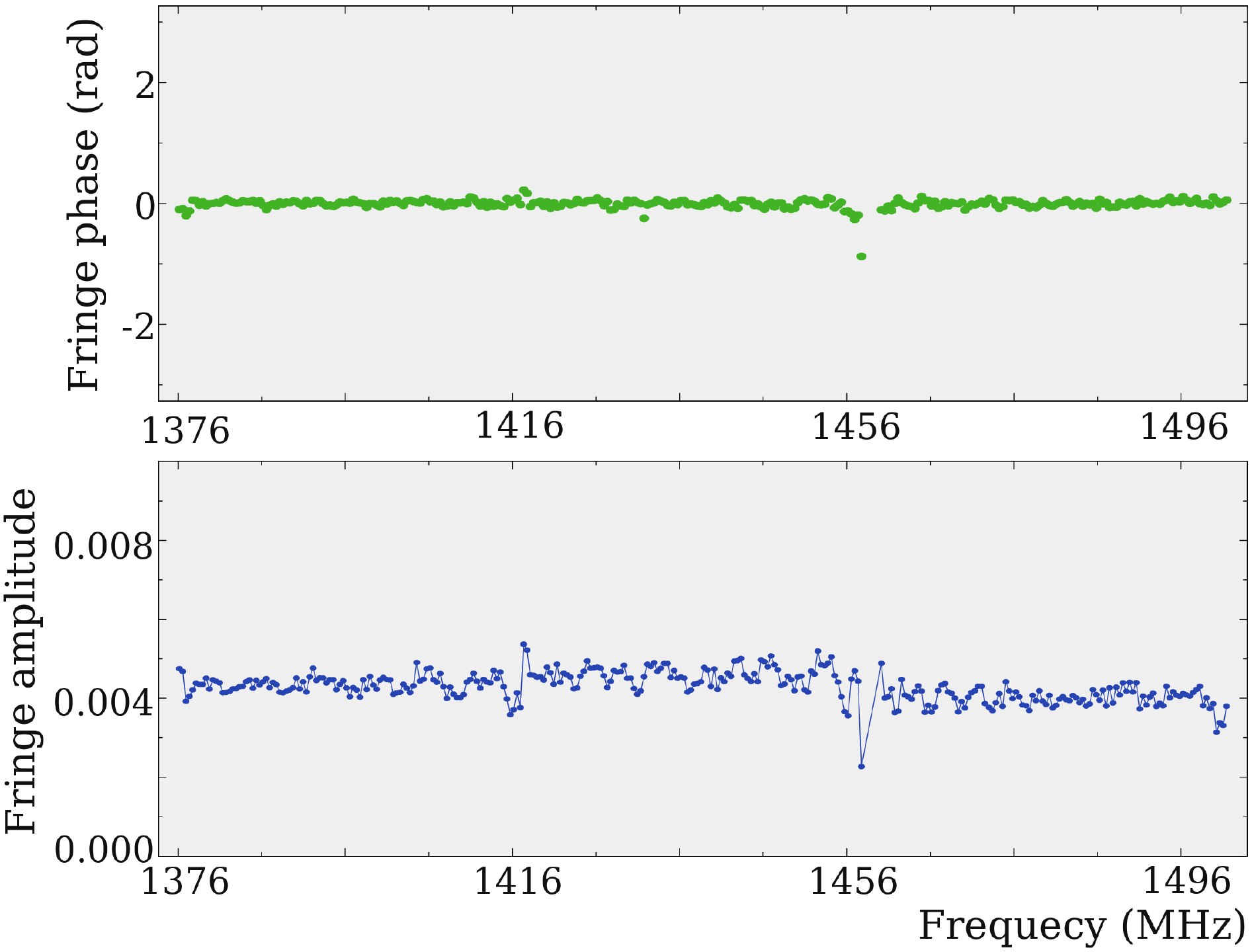}
  \hspace{0.01\textwidth}
  \includegraphics[width=0.487\textwidth]{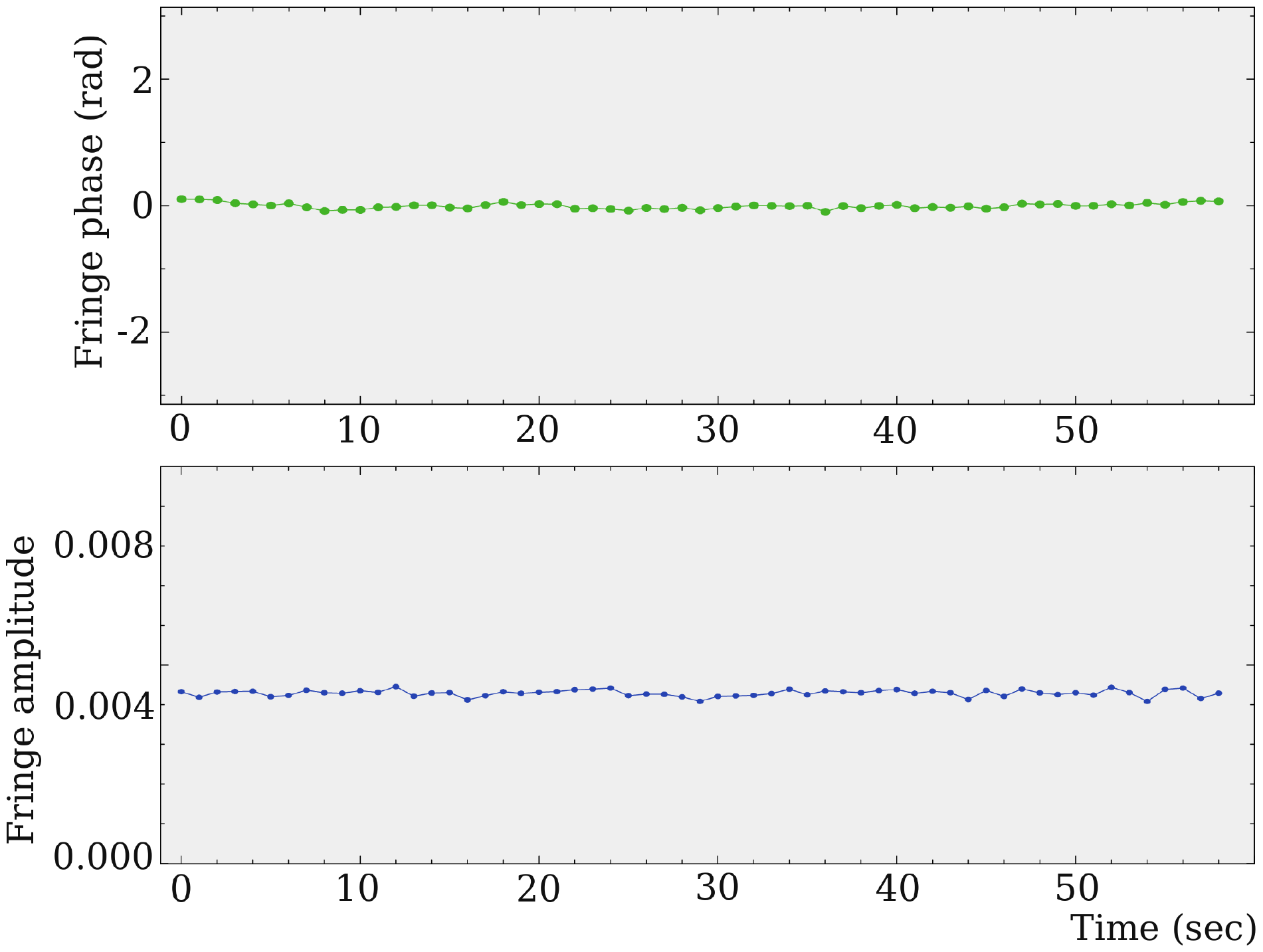}
  \caption{Fringe phase (upper) and normalized fringe 
           amplitude of Cyg-A at a 90~m long baseline 
           {\sc hrtr}/{\sc fd-vlba}. Signal to noise ratio 
           (SNR) 387 was achieved for 60~seconds of integration 
           time. Left plot shows fringe phase and amplitude 
           versus frequency and right plot shows fringe 
           phase and amplitude versus time.
          }
  \label{f:friplo_gal}
  \par\vspace{-2ex}\par
\end{figure}

  The observing schedule included observations of seven of the brightest 
radiogalaxies and a number of GNSS satellites. We have detected all but one radiogalaxy 
at the short baseline with {\sc fd-vlba} and some sources at the 9~km long 
baseline. As expected, no detection was found between the two HRTR stations. 
Figure~\ref{f:friplo_gal} shows fringe phases and normalized fringe 
amplitudes of radiogalaxy Cyg-A located at a distance of 
$7.2 \cdot 10^{21}$~m. This goes well beyond (fourteen orders of 
magnitude!!) the intended use of the GNSS equipment. The interferometric 
fringes of radiogalaxies were stable over time, and integration could be 
extended up to 20~minutes without a noticeable degradation of fringe 
amplitude.

\begin{figure}[h]
  \includegraphics[width=0.483\textwidth]{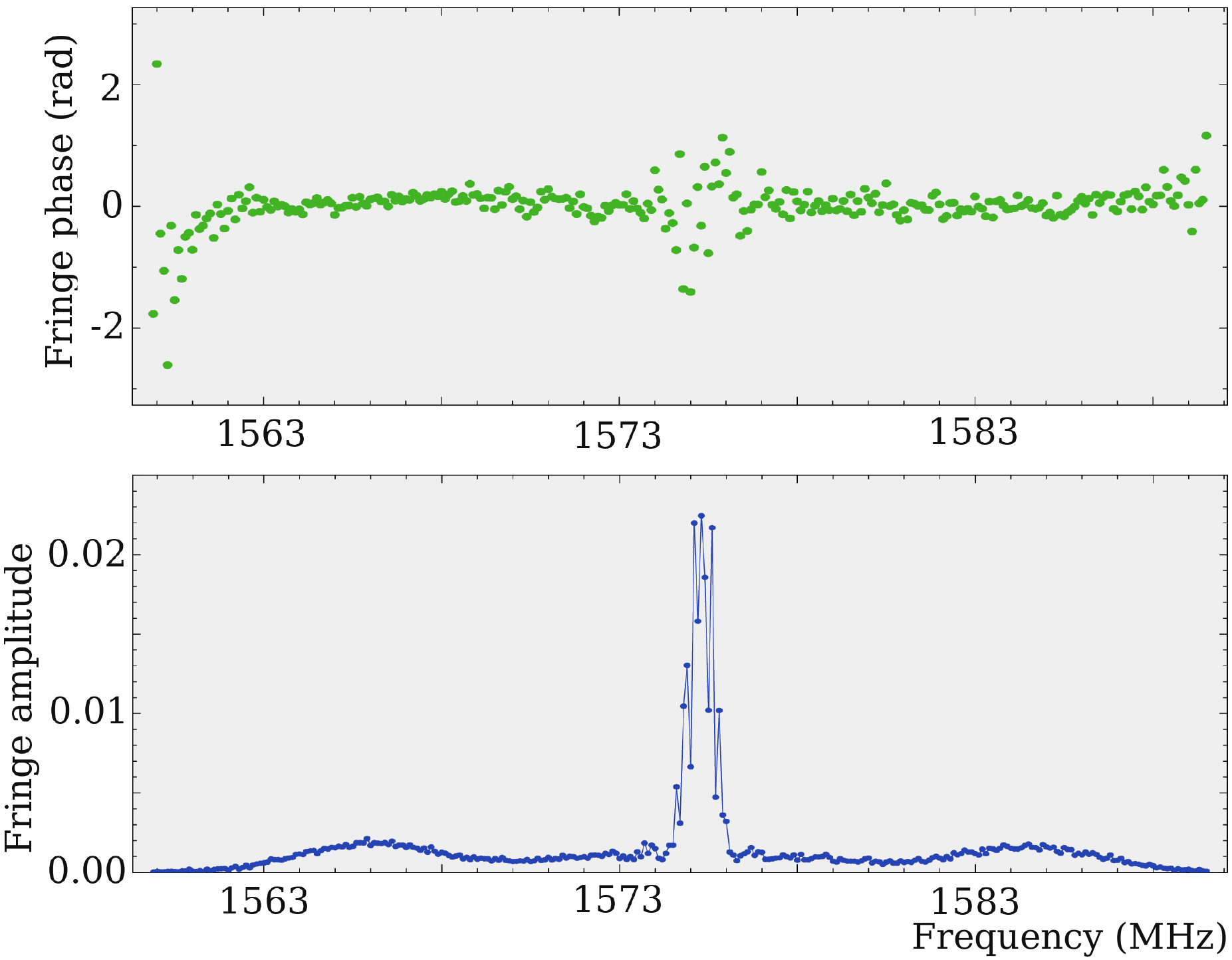}
  \hspace{0.01\textwidth}
  \includegraphics[width=0.483\textwidth]{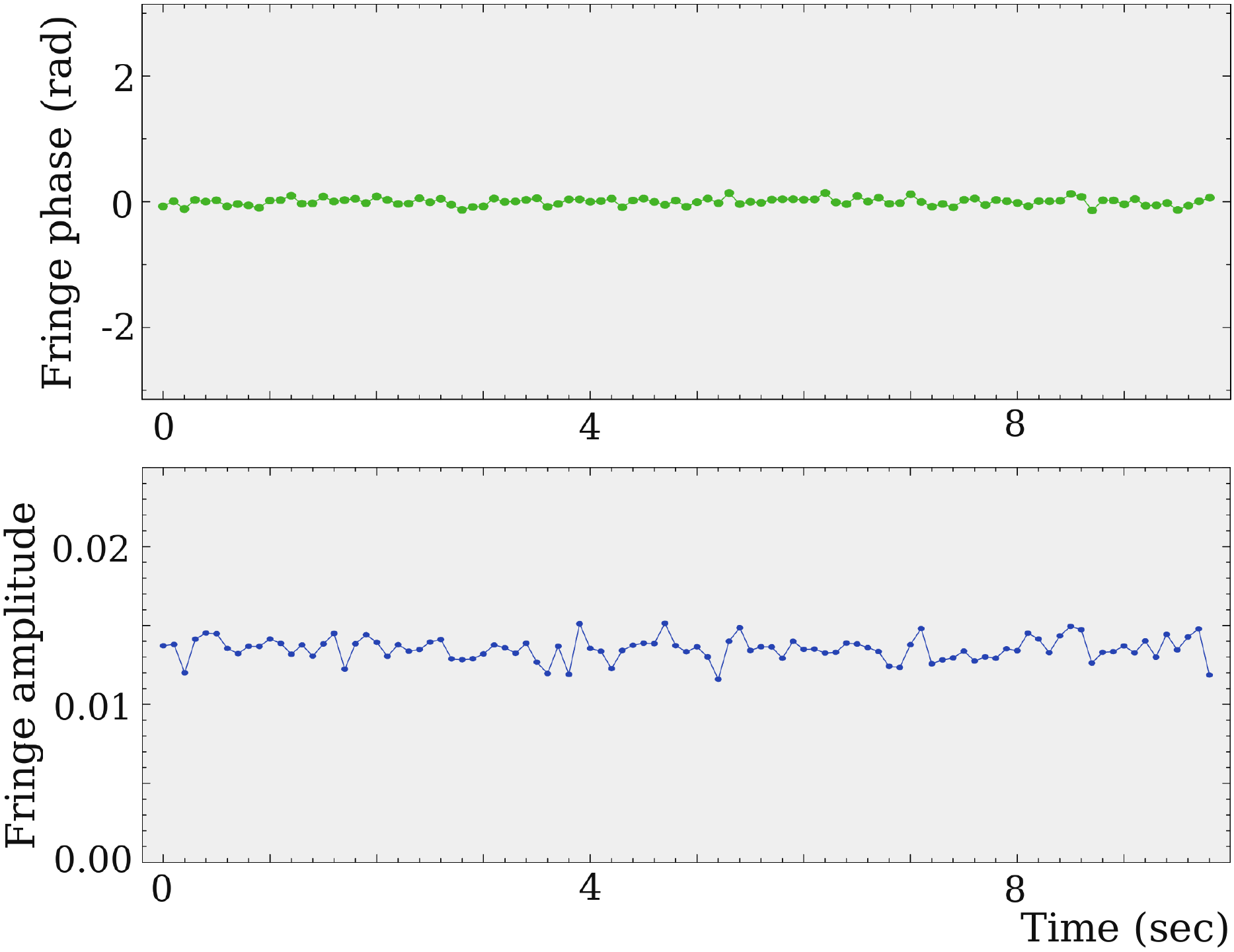}
  \caption{Fringe phase (upper) and normalized fringe 
           amplitude of a GPS satellite at a 90~m long 
           baseline {\sc hrtr}/{\sc fd-vlba}. SNR 138 was 
           achieved for 10~seconds of integration time.
           Left plot shows fringe phase and amplitude 
           versus frequency and right plot shows fringe 
           phase and amplitude versus time.
          }
  \label{f:friplo_gps}
  \par\vspace{-3ex}\par
\end{figure}

  Figure~\ref{f:friplo_gps} shows fringe plots of a GPS satellite
over 10~s integration time. The GPS signal was treated as a random 
noise in the model of our VLBI data analysis. The fringe amplitude
has a peak at the carrier frequency of 1575.42~MHz, emission
near 1~MHz of the peak due to the C/A signal, and a broad emission
due the binary offset carrier modulation of the M-code that 
has a detectable power within $\pm 15$~MHz of the carrier. This 
allows us to compute group delay over the total bandwidth of 
$\sim\! 30$~MHz with a precision of 60--90~ps over 10~s.
This should be sufficient for resolving phase delay ambiguities 
with spacings of 635~ps. Interferometric responses have been 
detected at the 9~km long baselines as well. At this stage of 
the project we did not yet attempt to perform geodetic analysis.
                
\section{Discussion}

  A vector tie can be determined from both observations of natural 
sources, such as radiogalaxies, and from observations of navigational
satellites. The broadband GPS signal due to the modulation of the 
M-code has flux density around -200~dBW/m${}^2$/Hz \citep{r:tho19}, i.e.
$\sim\! 1$~MJy, while there are only 10~natural radio sources brighter 
that 0.02~MJy at 1.5~GHz, i.e. a factor of 50 fainter. The scarcity 
of very strong radiogalaxies makes preparation of a VLBI schedule 
optimized for geodesy difficult.

  Detection of a radio source requires observations with a sensitive 
VLBI antenna. {\sc fd-vlba} is dedicated for astronomy and its L-band receiver 
has a system equivalent flux density (SEFD) of 289~Jy. Compare with SEFD 
2500~Jy at the 2--3~GHz band at 12~m geodetic VLBI antennas. Since sensitivity 
of an interferometer is proportional to the square root of the product of SEFD 
of individual antennas, VLBI observations of a GPS satellite between a HRTR 
and a radiotelescope requires $50^2=2500$ times less sensitive radiotelescope
than observations of radiogalaxies. 
Figures~\ref{f:friplo_gal}--\ref{f:friplo_gps} seem to contradict that 
statement. It turned out {\sc fd-vlba} receiver worked in a saturated regime 
when observing a megajansky source, and fringe amplitude was strongly 
underestimated. Applying an additional attenuation is required to mitigate 
the problem. But in general, it is much easier to reduce the sensitivity 
than to increase it.

  The primary observable that will be used for determination of the tie
vector will be phase delay. Phase delay ambiguities can be resolved if
group delay can be determined with an accuracy 1/4 of the phase ambiguity
spacing or better and short-term systematic differences between phase 
and group delays have a scatter not exceeding that number. Since precision
of group delay is reciprocal to the bandwidth of a detected signal, 
observations of navigational satellites with a relatively broad spectrum,
such as GPS and Galileo, is preferable.

  Determination of a tie vector between a HRTR and a geodetic VLBI antenna
is interesting, but not exactly what is needed. We need to determine a tie
vector between a VLBI antenna and a permanent GNSS antenna. There are 
two possible solutions. First, one can make HRTR a permanent GNSS site.
It is expected that in 2023--2024, permanent HRTRs will be installed within
a hundred meters of each of the ten VLBA antennas. Second, one can
install a HRTR on a temporary monument and determine a tie vector
HRTR/radiotelescope using VLBI and a tie vector between the HRTR and
the existing GNSS antenna by processing GNSS signals from both antennas
using double differences. Processing differential GNSS data at baselines 
of $\sim\!\! 100$~m long provides a millimeter level accuracy owing to 
cancellation of the atmospheric contribution. Then combining 
the VLBI/HRTR and HRTR/GNSS tie vectors, we get a VLBI/GNSS tie. 

  Radiotelescopes used for radio astronomy often have receivers in 
1.2--1.8~GHz ranges, although very few instruments have the ability to 
simultaneously record within a band of 1.15--1.65~GHz that covers
navigation signals L1, L2, and L5. Radiotelescopes dedicated for geodesy 
usually cannot receive emission below 2~GHz because of strong high-pass 
filters installed to mitigate the impact of radio interference. Some new 
generation broad-band VLBI Geodetic Observing System (VGOS) radiotelescopes
have a low cutoff frequency as high as 3.0~GHz because broadband receivers 
are much more susceptible to radio interference. A solution is to equip 
existing geodetic radio telescopes with auxiliary receivers operating 
in the 1--2~GHz range dedicated to VLBI observations of navigational 
satellites. Since the GNSS signal is so strong, such receivers do not 
require cooling. The navigational receiver can be installed alongside
the main receiver, and the signal can be directed to it either with 
a deployable mirror or with a dichroic plate.

\section{Summary}

   We propose a novel concept of GNSS/VLBI tie measurements based on a
microwave technique. We essentially transform a GNSS antenna into an
element of the VLBI network. This method allows us to estimate the tie
vector between the VLBI and GNSS reference points directly using the
microwave technique without the need to determine the position offsets
of the microwave reference points with respect to markers accessible to local 
surveys.

  We expect that the application of this method will have a profound impact
because we expect this method will be bias-free. As a result, vector
tie repeatability can be used as a measure of the accuracy of tie vector
determination. Knowing errors of tie vectors will enable us to close 
the budget of the differences of the VLBI reference points reduced to 
the GNSS reference points and make an inference about whether these 
differences are statistically significant or not.

  We ran three 3 hour long observing sessions between the {\sc fd-vlba} 
radiotelescope and a high rate GNSS receiver co-located within 90~m. 
We have detected fringes from both natural extragalactic radio sources 
and GPS satellites with  the SNR well above 100. Thus, we have 
demonstrated that technical problems related to GNSS/radiotelescope 
VLBI can be solved. Future work will be focused on the determination 
of vector ties using this technique.

\bibliography{tomm}

\end{document}